\documentclass[12pt]{article}

\usepackage{hyperref}
\title{Self-assembly of Microcapsules via Colloidal Bond Hybridization and Anisotropy}
\author{Chris H.J. Evers, Jurriaan A. Luiken, Peter G. Bolhuis\\ and Willem K. Kegel} 
\date{Please cite as: \href{http://dx.doi.org/10.1038/nature17956}{Evers \textit{et al. Nature} \textbf{534} 364--368 (2016)}}

\usepackage{textcomp}
\usepackage[geometry]{ifsym}
\usepackage{xcolor}
\usepackage{amsmath}

\usepackage{graphicx}

\usepackage{natbib}
\newcommand{\JournalTitle}[1]{\textit{#1}}

\definecolor{xBrown}{cmyk}{0,.10,1,.76}
\definecolor{xBlue}{cmyk}{1,.81,0,.53}
\definecolor{xRed}{cmyk}{0,1,1,.29}
\definecolor{xGreen}{cmyk}{.86,0,1,.06}
\definecolor{xLightred}{cmyk}{0,.82,.82,0}

\begin{document}
\maketitle
\begin{abstract}
\textbf{
Particles with directional interactions 
  are 
    promising building blocks 
    for new functional materials
  and 
    may serve as
    models for biological structures\cite{zhenli2004patchySA,glotzer2007anisotropy,yi2013patchySAR}.
Mutually attractive nanoparticles
  that are deformable due to flexible surface groups,
  for example,
  may spontaneously order themselves into strings, sheets and large vesicles\cite{akcora2009anisotropicSAPolymerNP,nikolic2009amphiphilicNPVesicles,luiken2013anisotropicPolymerNP}. 
Furthermore,
  anisotropic colloids 
  with attractive patches
  can self-assemble into 
  open lattices
  and colloidal equivalents of molecules and micelles\cite{chen2011kagomeSA,wang2012colloidsValence,kraft2012surfaceRoughnessSelfAssembly}.
However,
  model systems 
  that combine mutual attraction, anisotropy, and deformability
  have\textemdash 
  to the best of our knowledge\textemdash 
  not been realized.
Here,
  we synthesize colloidal particles
  that combine these three characteristics
  and obtain self-assembled microcapsules.
We propose that
  mutual attraction and deformability
  induce
  directional interactions via
  colloidal bond hybridization.
Our particles
  contain
  both mutually attractive and repulsive surface groups
  that are flexible.
Analogous to the simplest chemical bond,
  where two isotropic orbitals hybridize into the molecular orbital of H$_2$,
  these flexible groups redistribute upon binding.
Via colloidal bond hybridization,
  isotropic spheres
  self-assemble into planar monolayers,
  while anisotropic snowman-like particles
  self-assemble into hollow monolayer microcapsules.
A modest change of the building blocks
  thus results in a significant leap in the complexity of the self-assembled structures. 
In other words, 
  these relatively simple building blocks 
  self-assemble into dramatically more complex structures
  than similar particles that are isotropic or non-deformable.
} 
\end{abstract}

For self-assembly of nanoparticles,
  deformability and mutual attraction
  have recently been combined
  by grafting flexible polymers onto the surface of mutually attractive particles.
This 
  results in 
  isotropic clusters  \cite{larsonSmith2011saNPControlledStericInteractions},
  and 
  self-assembled strings, sheets, and large vesicles\cite{akcora2009anisotropicSAPolymerNP,nikolic2009amphiphilicNPVesicles}.
For micrometre-sized colloids,
  on the other hand,
  coupling mutual attraction and anisotropy 
  leads to \textit{patchy particles}.
Attractive domains, or \textit{patches},
  have induced self-assembly
  into open lattices
  and colloidal equivalents of molecules and micelles\cite{chen2011kagomeSA,wang2012colloidsValence,kraft2012surfaceRoughnessSelfAssembly}.
Here,
  we 
  combine the three properties mutual attraction, anisotropy and deformability, 
  by synthesizing
  snowman-like particles 
  that consist of a deformable core and a non-deformable lobe or \textit{protrusion}.
In the first part of this letter,
  mutual attraction is combined with deformability,
  resulting in
  anisotropic or \textit{directional} interactions
  as
  flexible surface groups redistribute upon binding (Fig. \ref{fgr:Isotropic}e).
This process 
  is analogous to bond hybridization in quantum chemistry.
When two hydrogen atoms bind and form H$_2$,
  for example,
  the electrons around each atom redistribute, 
  i.e. two isotropic orbitals 
  hybridize into the molecular orbitals of H$_2$.
Similarly,
  when mutually attractive, deformable particles bind,
  flexible surface groups redistribute,
  resulting in directional interactions.
We refer to this effect 
  as colloidal bond hybridization.
We observe 
  fundamentally new behaviour 
  upon combining colloidal bond hybridization with anisotropy,
  i.e. for particles that are mutual attractive and deformable as well as anisotropic.
These snowman-like particles
  self-assemble
  into microcapsules,
  and form spherical cavities
  at high particle concentrations.
We hypothesize that 
  mutual attraction, anisotropy and deformability 
  are sufficient to stabilize
  curved structures, 
  and we make this hypothesis more plausible with computer simulations.

\vspace{1cm}
We create
  isotropic as well as anisotropic building blocks
  that are mutually attractive and deformable.
Before 
  discussing the more complex anisotropic particles,
  we consider the basic principles of colloidal bond hybridization
  using mutually attractive, isotropic, deformable particles.
These 
  poly(styrene-\textit{co}-acrylic acid) spheres
  are synthesized by 
  copolymerization in water ({Fig. \ref{fgr:Isotropic}a--b),
  and  
  acrylic acid and styrene
  are incorporated at different stages in the polymerization process\cite{wang2002styreneAcrylicAcid,hu2009snowman-likePolystyrene}.
Hence,
  the particles consist of 
  a hydrophobic polystyrene-rich core
  and a hydrophilic poly(acrylic acid)-rich brush.
The particles 
  are mutually attractive
  as hydrophobic polystyrene groups 
  are present both in the interior of the particles
  as well as in the poly(acrylic acid)-rich brush.
Furthermore,
  dynamic light scattering
  shows that the poly(acrylic acid)-rich brush
  can 
  rearrange on the order of 0.1 \textmu m,
  rendering the particles
  deformable (Extended Data Fig. \ref{exfgr:DLS}).

\begin{figure}
	\centering
	\includegraphics{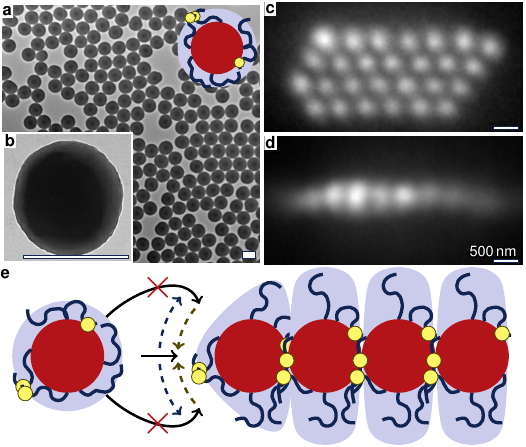}
  \caption{
\textit{Self-assembled planar monolayers.} 
Mutually attractive, isotropic, deformable particles (a--b, transmission electron microscopy) consist of a polystyrene-rich core (red), a deformable poly(acrylic-acid)-rich brush (blue) and mutually attractive moieties (yellow). In water, the  particles self-assemble into planar hexagonal monolayers (c--d, top and side view, reflected light microscopy). 
In colloidal bond hybridization (e), surface groups of deformable particles redistribute upon binding; mutually attractive moieties (yellow) move towards the contact area, while hydrophilic chains (blue) move into the solution.
}
  \label{fgr:Isotropic}
\end{figure}

Mutually attractive, isotropic, deformable  particles
  self-assemble into planar monolayers
  in water (Fig. \ref{fgr:Isotropic}c--d).
The monolayer sheets
  are hexagonally ordered,
  and move freely in the solution (Supplementary Video 1).
We hypothesize 
  that a colloidal equivalent of bond hybridization
  drives the formation of monolayers.
The polymer brush contains
  hydrophobic styrene groups
  as well as hydrophilic acrylic acid groups (Fig. \ref{fgr:Isotropic}e yellow and blue).
Attraction between the hydrophobic groups
  promotes compact structures,
  while excluded volume effects of the hydrophilic parts
  favour unbound particles.
To accommodate both effects,
  the polymer brush
  rearranges 
  upon binding:
  hydrophobic parts interact in-plane,
  while hydrophilic parts 
  expand out-of-plane.
Consequently,
  directional interactions
  are induced
  and
  planar monolayers are formed (Fig. \ref{fgr:Isotropic}e).

This segregation process
  is similar to
  phase segregation 
  in self-assembly of block copolymers\cite{bates1991polymerPolymerPhaseBehavior}.
In our system,
  however,
  copolymers are anchored to the surface of \textmu m-sized particles.
Consequently,
  molecular segregation of the polymers 
  induces directional interactions on the colloidal length scale.
Our observations are also in line 
  with results for polymer-grafted nanoparticles
  that are mutually attractive, isotropic and deformable\cite{akcora2009anisotropicSAPolymerNP,nikolic2009amphiphilicNPVesicles,luiken2013anisotropicPolymerNP,asai2015polymerGraftedNPpatchyParticles}.
In our system,
  however,
  directional interactions are induced
  for particles that are two orders of magnitude larger than in previous work.
Finally, 
  DNA coated colloids
  can also form crystalline monolayers\cite{geerts2010flyingCarpets},
  but for these particles,
  a functionalized surface
  induces directional interactions.

\vspace{1cm}
In the remainder of this paper,
  we combine colloidal bond hybridization with anisotropy,
  which results in fundamentally new behaviour.
Anisotropic building blocks 
  are synthesized
  by growing a rigid protrusion onto the deformable spheres of Fig. \ref{fgr:Isotropic} (Extended Data Fig. \ref{exfgr:Synthesis}).  
The second lobe
  is grown
  by swelling with additional styrene\cite{sheu1990nonsphericalLatex,mock2006synthesisAnisotropicNP,kraft2009colloidalMoleculesBondAngles}.
Furthermore,
  we increase the attraction
  between the deformable lobes by functionalizing the poly(acrylic acid)-rich brush with hydrophobic groups\cite{MarchandBrynaert1995surfacefunctionalization} (Extended Data Fig. \ref{exfgr:Fluorescence}).
Next,
  the particles are washed by centrifugation,
  which is a crucial step as we will see later.
Finally, we obtain 
  snowman-like particles 
  that consist of a deformable lobe
  and a non-deformable lobe (Fig. \ref{fgr:Microcapsules}a).

\begin{figure}
	\centering
	\includegraphics{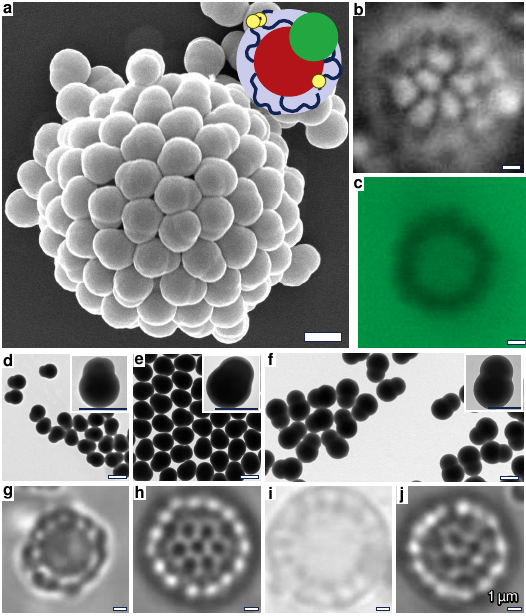}
  \caption{
\textit{Self-assembled monolayer microcapsules.} 
Deformable, anisotropic, mutually attractive particles consist of a core (a, red) with hydrophilic (blue) and hydrophobic (yellow) moieties and a rigid protrusion (green), 
and self-assemble into microcapsules (a, scanning electron microscopy after sintering). In solution, the particles align tangentially (b, reflected light microscopy), and the microcapsules are water-filled (c, confocal fluorescence microscopy with dyed water phase). Particles are synthesized with large lobes of diameters 445, 540 and 626 nm (d--f, transmission electron microscopy). For each size (g--i, bright field microscopy), and upon functionalization with both fluoresceinamine (g--i) and \textit{tert}-butylamine (j, bright field microscopy), self-assembled microcapsules are found.
}
  \label{fgr:Microcapsules}
\end{figure}

These mutually attractive, anisotropic, deformable particles
  self-assemble into
  monolayer microcapsules (Fig. \ref{fgr:Microcapsules}).
The microcapsules
  can be observed 
  after sintering or freeze drying with scanning electron microscopy (Fig. \ref{fgr:Microcapsules}a, Extended Data Fig. \ref{exfgr:SEM}).
Due to the relatively large size of the particles, 
  however,
  we can even study structures in solution
  using optical microscopy (Fig. \ref{fgr:Microcapsules}g--j).
The microcapsules
  consist of 
  a particle monolayer 
  while the interior is water-filled (Fig. \ref{fgr:Microcapsules}c, Supplementary Video 3).
Furthermore, 
  most particles align tangentially to the surface of the microcapsules, with the protrusions pointing either slightly inwards or slightly outwards (Fig. \ref{fgr:Microcapsules}a--b, Extended Data Fig. \ref{exfgr:SEM}). 
    
For particles with a large lobe of 0.540 \textmu m in diameter,
  the mean diameter of the microcapsules  
  is $3.7\pm0.8$ \textmu m
  corresponding to about $10^2$ particles per microcapsule. 
Most particles have six nearest neighbours,
  but pentagons occur frequently
  as expected from the Euler characteristic
  of a sphere (Fig. \ref{fgr:Microcapsules}a, Supplementary Video 2).
The structure of the microcapsules,
  however,
  has no overall icosahedral symmetry.
Furthermore,  
  excess styrene is removed before the microcapsules are formed,
  so unlike colloidosomes\textemdash 
  that are formed on emulsion droplets\cite{dinsmore2002colloidosomes}\textemdash 
  no template is involved.

By systematically varying the complexity of the particles,
  we identify 
  that in our system,
  mutual attraction, anisotropy and deformability
  are required
  for self-assembly into microcapsules (Extended Data Fig. \ref{exfgr:Capillary}a--d). 
First, before functionalization with mutually attractive groups,
  no microcapsules are found for any of the anisotropic, deformable particles in Fig. \ref{fgr:Microcapsules}d--f,
  showing the importance of mutual attraction.
Secondly,
  for hydrophobically functionalized, deformable, but isotropic spheres,
  also no microcapsules are observed.
Finally, 
  for functionalized, anisotropic, but non-deformable snowman-like particles,
  no microcapsules are observed either. 
These three characteristics
  seem sufficient 
  to induce self-assembly into microcapsules,
  and are relatively easy to experimentally implement.
In contrast,
  particles with four orthogonally attractive patches,
  that have previously been predicted to induce self-assembly into microcapsules,
  have not been experimentally realized yet\cite{chen2007coneShapedSelfAssembly}. 
Moreover,
  the self-assembling tendency 
  is robust,
  as both monolayer microcapsules and planar monolayers are found 
  for snowman-like particles with large lobes of diameters ranging from 445 to 626 nm, 
  and for hydrophobic functionalization with either \textit{tert}-butylamine or fluoresceinamine  (Fig. \ref{fgr:Microcapsules}d--j, Extended Data Fig. \ref{exfgr:Sheets}).

\vspace{1cm}

At high particle concentrations, 
  mutually attractive, anisotropic, deformable particles
  form curved hollow structures or \textit{cavities},
  and these 
  are likely intermediates in the formation of microcapsules.
The tendency to form cavities 
  is robust,
  as similar structures are observed
  at five different experimental conditions.
The first condition
  is at the edge of an evaporating droplet on a glass slide (Fig. \ref{fgr:Cavities}a--e).
Evaporation of a droplet of particles in water
  causes 
  a particle flow towards the
  glass/water/air contact line 
  that is known as the coffee stain effect\cite{deegan1997coffeeStainCapillaryFlow}.
For our particles,
  hemispherical cavities are spontaneously formed
  in the resulting dense layer near the contact line (Fig. \ref{fgr:Cavities}b--e, Extended Data Fig. \ref{exfgr:ContactLine}a--d).
The second condition
  is for a droplet
  that is confined between two parallel glass slides (Fig. \ref{fgr:Cavities}f--j, Supplementary Video 4).
Again,
  a dense layer is formed at the contact line, 
  but now 
  the layer is two-dimensional with 
  circular cavities.
The third condition is 
  at particle volume fractions of $\sim$0.2,
  where a highly fluctuating `cavity phase'
  is formed (Fig. \ref{fgr:Cavities}k--n, Supplementary Video 5).
In this phase,
  we observe
  coexisting regions on the order of 1--10 \textmu m 
  with either high particle concentrations or virtually no particles,
  i.e. dense curved structures around cavities.
Fourthly,
  upon centrifuging particles in a capillary,
  cavities are observed in the sediment (Fig. \ref{fgr:Cavities}q--s and Extended Data Fig. \ref{exfgr:Sediment}a).
Finally,
  upon diluting the sediment,
  again the formation of circular cavities is observed (Fig. \ref{fgr:Cavities}t--v,  Supplementary Video 6).

\begin{figure}
	\centering
	\includegraphics{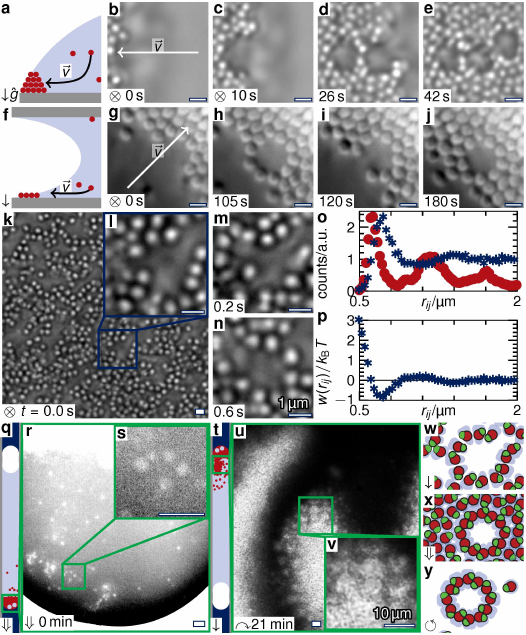}
  \caption{
\textit{Cavity formation upon densification.} Cavities are observed at the edge of an evaporating droplet (a--e, multilayer, f--j, monolayer) in a \textit{cavity phase} (k--n), after centrifugation in a thin cell (q--s, bright regions), and upon turning the cell upside down (t--v). The histograms of the interparticle distance, $r_{ij}$ (o) peak at 0.72 \textmu m (cavity phase, blue), and at 0.62 \textmu m (microcapsules, red), and  the effective pair potential, $w(r_{ij})$ (p) has a minimum of $-0.8\,k_\textup{B}T$. Particles form a cavity phase upon densification (w), and are pushed into close contact upon centrifugation (x). We propose that after redispersion (y), particles that surrounded cavities are found as microcapsules. The arrows indicate the flow, $\vec{v}$, and the gravitational field $\hat{g}$.   
}
  \label{fgr:Cavities}
\end{figure}

At all five conditions,
  the diameters of the cavities 
  are 
  comparable to the diameters of the microcapsules.
Furthermore,
  by systematically varying the complexity of the building blocks,
  we conclude that\textemdash
  as for the formation of microcapsules\textemdash
  mutually attraction, anisotropy and deformability
  all greatly influence 
  the formation of cavities (Extended Data Fig. \ref{exfgr:Capillary}).
For isotropic particles,
  we proposed that mutual attraction and deformability 
  induce the observed self-assembly into planar monolayers
  by colloidal bond hybridization. 
Based on the above observations, 
  we hypothesize that 
  adding anisotropy to mutually attractive and deformable colloids
  induces a shift
  from planar to curved structures,
  resulting in microcapsules and cavities.

The formation of microcapsules
  is a rare event as 
  only 1 in every $10^4$ particles 
  ends up in a microcapsule.
This can be ascribed
  to the specific orientation of many particles
  required for the formation of microcapsules,
  and the initially weak attractive interactions.
The latter 
  becomes apparent
  as the minimum in the effective pair potential is comparable to the thermal energy, $k_\textup{B}T$ (Fig. \ref{fgr:Cavities}p),
  and particles do not form lasting clusters upon collision (Supplementary Video 5).
Furthermore,
  for particles in such non-lasting clusters,
  the centre-to-centre distance distribution 
  peaks at a 0.1 \textmu m larger distance
  than for particles in microcapsules (Fig. \ref{fgr:Cavities}o).
The difference in the centre-to-centre distance 
  can be attributed to 
  the hydrophilic chains with an estimated length of about 0.1 \textmu m.
These chains 
  need to
  move out of the binding site 
  upon bond formation (Fig. \ref{fgr:Isotropic}e).

We propose
  that both the specific orientation
  and the formation of lasting bonds
  are induced by centrifugation
  with cavities as intermediates.
The synthesis contains several centrifugation steps 
  and centrifugation
  induces the formation of spherical cavities with a similar size and shape as microcapsules (Fig. \ref{fgr:Cavities}q--s).
Centrifugation 
  thus aligns the particles in a specific microcapsule-like orientation.
Furthermore,
  upon centrifugation,
  particles are pushed close together.
This could push the hydrophilic chains out of the binding side,
  and induce the formation of irreversible bonds
  that arise from van der Waals forces between the colloids at close proximity.
While 
  we have no 
  direct, real-space proof 
  of the formation mechanism, 
  microcapsules could be formed as follows: 
  first,
  particles form a dense sediment 
  with cavities (Fig. \ref{fgr:Cavities}w);
  next, 
  centrifugation pushes particles closer together 
  and irreversible bonds are formed (Fig. \ref{fgr:Cavities}x);
  finally, after shaking,
  particles that surrounded cavities are found as microcapsules (Fig. \ref{fgr:Cavities}y).
\vspace{1cm}

We test the key hypothesis
  that combining 
  colloidal bond hybridization with anisotropy
  can stabilize curved monolayers
  using Monte Carlo simulations.
First,
  we develop a simple model 
  for mutually attractive, isotropic, deformable particles.
Next,
  we extend this model 
  with anisotropy.

Mutually attractive, isotropic, deformable particles
  are modelled 
  as central spheres
  with $f$ satellite spheres each (Fig. \ref{fgr:Sim}b).
The size ratio between the central and the satellite spheres 
  is $q$,
  and the latter
  are penetrable hard spheres
  that model the flexible surface groups.
Penetrable hard spheres
  can interpenetrate other satellite spheres,
  but have excluded volume interactions with the central spheres.
Mutual attraction 
  is captured 
  by a square well interaction between central spheres,
  and deformability 
  is incorporated 
  as the satellite spheres
  can freely move over the surface of the central sphere.

\begin{figure}
	\centering
	\includegraphics{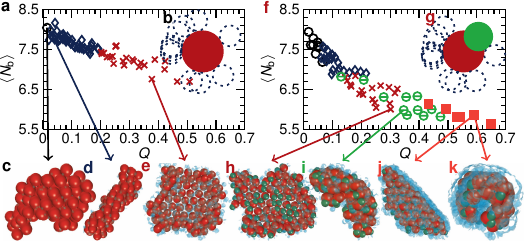}
  \caption{
\textit{Monte Carlo simulations.} b) Deformable spheres are modelled as attractive spheres (red) with $f$ mobile penetrable hard spheres (blue). a) The average number of bonds, $\langle N_\textup{b} \rangle$ decreases with the covered surface fraction $Q$, resulting in 
compact clusters (c,\SmallCircle), 
bilayers (d,\textcolor{xBlue}{\SmallDiamondshape}) 
and monolayers (e,\textcolor{xRed}{\SmallCross}). 
g) Snowman-like particles are modelled as deformable spheres with a rigid protrusion (green). f) Adding a protrusion induces a shift from 
flat monolayers (h,\textcolor{xRed}{\SmallCross})
to curved monolayers with in-plane protrusions (i,\textcolor{xGreen}{\rlap{\HBar}\SmallCircle}) 
and out-of-plane protrusions (j--k,\textcolor{xLightred}{\FilledSmallSquare})
at high $Q$-values.
}
  \label{fgr:Sim}
\end{figure}

To verify 
  if colloidal bond hybridization can induce directional interactions,
  i.e. if rearrangement of surface groups
  can stabilize monolayers,
  we start Monte Carlo simulations in a hexagonal planar configuration.
Different morphologies 
  are observed upon varying 
  the size and the number of satellite spheres (Extended Data Fig. \ref{exfgr:Morphology}b,i--k).
These two variables 
  can be combined in the covered surface fraction, $Q=\frac{fq^2}{4(1+q)^2}$.
Upon plotting
  the average number of bonds, $\langle N_\textup{b} \rangle$,
  as a function of $Q$, 
  all data collapse onto a single curve (Fig. \ref{fgr:Sim}a).
Furthermore,
  the transitions between different morphologies occur at well defined $Q$-values 
  showing that the covered surface fraction 
  dictates the morphology (Fig. \ref{fgr:Sim}a, Extended Data Fig. \ref{exfgr:Morphology}b).
If the covered surface fraction is small,
  the particles reorganize into compact aggregates with many bonds
  to maximize the attractive interaction between the central spheres (Fig. \ref{fgr:Sim}c).
Upon increasing the size or number of satellite spheres,
  however,
  they redistribute out-of-plane 
  and mechanically stable bilayers and monolayers are observed (Fig. \ref{fgr:Sim}d--e).
Colloidal bond hybridization
  can thus induce directional interactions
  and stabilize monolayers for isotropic particles,
  which agrees well with the results in Fig. \ref{fgr:Isotropic}.

Starting with unbound particles (Extended Data Fig. \ref{exfgr:Morphology}a,c--h),
  we qualitatively reproduce 
  simulation results 
  that have been obtained using a more detailed model with tethered chains\cite{akcora2009anisotropicSAPolymerNP}.
Hence, 
  we conclude that, albeit simple, our model
  captures the main ingredients 
  that induce directional interactions:
  mutual attraction and deformability.
As our model does not contain any molecular details,
  we expect this behaviour
  to be generic.
This expectation is in line with
  experimental systems 
  of polymer-grafted nanoparticles
  where both mutual attraction and deformability can be identified
  and directional interactions are induced\cite{akcora2009anisotropicSAPolymerNP,nikolic2009amphiphilicNPVesicles}.

We model mutually attractive, anisotropic, deformable particles
  by adding 
  a rigid sphere
  to the deformable sphere (Fig. \ref{fgr:Sim}g, Extended Data Fig. \ref{exfgr:MorphologySnowmen}).
The rigid sphere
  models the polystyrene protrusion
  in the most primitive way, 
  and its hydrophobicity 
  is captured
  by a short-ranged attraction 
  with other rigid and central spheres.
Similarly as for isotropic particles,
  the values for the average number of bonds
  collapse on a single curve 
  as a function of the covered surface fraction (Fig. \ref{fgr:Sim}f, Eq. \ref{eq:csfSnowmen}),
  and 
  planar monolayers are stable
  at moderate values for the covered surface fraction (Fig. \ref{fgr:Sim}h).
Upon increasing the size and the number of satellite spheres further,
  however,
  curved monolayers 
  with protrusions pointing just out-of-plane (Fig. \ref{fgr:Sim}i)
  are observed,
  while 
  for large and many satellite spheres,
  all protrusions point inwards (Fig. \ref{fgr:Sim}j--k).

We conclude that,
  as hypothesized,
  1) colloidal bond hybridization 
  can stabilize monolayers,
  and 
  2) anisotropy induces
  a shift from planar to curved structures.
Furthermore,
  hemispherical monolayers with in-plane protrusions (Fig. \ref{fgr:Sim}i, Supplementary Video 7)
  resemble segments of the experimentally observed microcapsules.

\vspace{1cm}
In this paper,
  we combined
  mutual attraction, anisotropy and deformability
  in colloidal model particles.
On the one hand,
  mutual attraction and deformability
  cause surface groups to rearrange upon binding,
  which is a colloidal equivalent of bond hybridization;
  on the other hand,
  anisotropy induces curvature.
These three characteristics 
  are most likely sufficient
  to induce self-assembly into microcapsules,
  a process that\textemdash
  to the best of our knowledge\textemdash
  had not been realized before in a colloidal model system. 
We note that,
  while details are different,
  in the building blocks of viruses,
  mutual attraction, deformability and anisotropy can also be identified\cite{berg2002biochemistry,tompa2008proteinFuzzyComplexes,freund2008structuralRearrangementsHepatitis},
  suggesting that these characteristics 
  could be 
  important in the assembly of virus microcapsules as well. 
The mechanism
  we find
  is fundamentally different
  from previous work
  where directional interactions are induced 
  by rigid patches\cite{chen2011kagomeSA,wang2012colloidsValence,kraft2012surfaceRoughnessSelfAssembly},
  structural rearrangements upon changing the solvent\cite{groschel2013softPatchyNPSA},
  or electric fields\cite{crassous2014fieldInducedEllipsoidAssemblyMicrotubules}.
Quantification of 
  attraction strength, anisotropy, number of particle lobes and deformability 
  lead to a large parameter space that remains to be systematically explored.
Additionally,
  more theoretical work is needed
  to predict
  self-assembled structures
  from these properties,
  for which work of Asai \textit{et al.} could be extended\cite{asai2015polymerGraftedNPpatchyParticles}.
Independently controlling these properties,
  however,
  seems impossible for proteins;
  colloidal particles,
  on the other hand,
  are promising building blocks to 
  address this challenge.

\clearpage

\paragraph{Supplementary Information} is available in the online version of the paper.
\paragraph{Acknowledgements}
The authors thank Bas van Ravensteijn for providing non-deformable, fluorescein functionalized snowman-like particles, Sonja Castillo for taking the scanning electron microscopy images and Hans Meeldijk and Chris Schneijdenberg for help with freeze drying and transmission electron microscopy.
This work is part of the research programmes VICI 700.58.442 and TOP-GO 700.10.355, which are financed by the Netherlands Organization for Scientific Research (NWO). Alfons van Blaaderen and Marjolein Dijkstra are thanked for useful discussion, and Mathijs de Jong is thanked for critically reading the manuscript. 
\paragraph{Author Contributions}
All authors designed the research; C.E. synthesised the particles and analysed the self-assembled structures; J.L. performed and analysed the Monte Carlo simulations. W.K. and P.B. supervised the project. All authors discussed the results and implications and wrote the paper.
\paragraph{Author Information}
Reprints and permissions information is available at www.nature.com/reprints. The authors declare no competing financial interests. Correspondence and requests for materials should be addressed to C.E. (c.h.j.evers@uu.nl) or W.K. (w.k.kegel@uu.nl).    

\clearpage
\section*{Methods}
\paragraph{Chemicals}   
Unless stated otherwise,
  the following chemicals 
  were used as received:
  acrylic acid (AA, 99\%),
  aluminum oxide (Al$_2$O$_3$, puriss., $\geq$98\%),
  \textit{tert}-butylamine (tBA, $\geq$99.5\%),  divinyl benzene (DVB, 55\%, mixture of isomers),
  fluorescein sodium salt (Fl, F6377),
  fluoresceinamine (FlA, mixture of isomers, $\geq$75\%), 
  \textit{N}-(3-Dimethylaminopropyl)-\textit{N}'-ethylcarbodiimide hydrochloride (EDC, purum, $\geq$98.0\%),
  2-(N-Morpholino)\-ethanesulfonic acid (MES,$\geq$99\%),
  sodium phosphate dibasic (Na$_2$HPO$_4$, BioXtra, $\geq$99\%),
  sodium phosphate monobasic dihydrate (NaH$_2$PO$_4\cdot2$H$_2$O, BioUltra, $\geq$99\%),
  and styrene (St, ReagentPlus, $\geq$99\%),
  were obtained from Sigma-Aldrich or its subsidiaries;
2,2'-azobis(2-methylpropionitrile) (AIBN, 98\%),
  potassium chloride (KCl, p.a.), 
  and potassium persulfate (KPS, reagent ACS, 99+\%)
  were obtained from Acros Organics;
hydroquinone (puriss, $\geq$99.5\%)
  was obtained from  Riedel-de Ha\"en;
glycerol (Ph Eur)
  was obtained from Bufa;
ethanol (100\%)
  was obtained from Interchema;
potassium hydroxide (KOH) 
  was obtained from Emsure;
hydrochloric acid (HCl, 37\%) 
  was obtained from Merck;
  and
Millipore water (MQ)
  was obtained with a Synergy water purification system.

\paragraph{Synthesis}   
\label{sup:synthesis}
The synthesis is outlined in Extended Data Fig. \ref{exfgr:Synthesis}a and
  involved
  (i) emulsifier-free polymerization of 
  cross-linked poly(styrene-\textit{co}-acrylic acid) (CPSAA) spheres,
  (ii) protrusion formation by swelling with styrene, heating and polymerizing 
  and (iii) covalently linking hydrophobic moieties to the carboxylic groups.

Cross-linked poly(styrene-\textit{co}-acrylic acid)  spheres
  of $0.530\pm0.014$ \textmu m in diameter
  are prepared by emulsifier-free polymerization of styrene, acrylic acid and divinyl benzene
  based on Wang and Pang and Hu \textit{et al.}
  \cite{wang2002styreneAcrylicAcid,hu2009snowman-likePolystyrene}.  
90 ml MQ, 
  11 ml} St 
  passed over an Al$_2$O$_3$ column,
  761 \textmu l freshly opened AA,
  and 55 \textmu l DVB 
  were added to a 250 ml three-neck round-bottom flask.
The flask 
  was constantly and vigorously stirred with a glass stirrer 
  under nitrogen flow.
Quantitatively, 
  0.05 g KPS 
  was dissolved as an initiator
  and added to the flask with 10 ml MQ.
After 15 minutes,
  the nitrogen inlet was raised above the liquid level,
  and after 15 more minutes,
  the flask 
  was immersed
  in a 70 \textdegree C oil bath
  to start the polymerization.
After 20 hours, 
  a milky-white dispersion
  was obtained.
Excess reactants
  were removed by centrifugation (Beckman Coulter Allegra X-12R).
Upon centrifugation, 
  particles settle at the bottom of the sample, 
  while unreacted chemicals were in the so-called supernatant. 
Three times,
  the dispersion was washed by centrifugation at $2.1\times10^3\,g$ and the supernatant was replaced by MQ. 

Cross-linked polystyrene (CPS) spheres
  are prepared in a similar method.
For these particles, 
  225 ml MQ, 
  23.5 ml St 
  and 0.7 ml DVB 
  were added to a 500 ml one-neck round-bottom flask.
The flask 
  was constantly and vigorously stirred with a PTFE coated stir bar,
  immersed in an 80 \textdegree C oil bath,
  and 0.78 g KPS in 37.5 ml MQ was added.
After 24 h a milky-white
  dispersion was obtained,
  which was washed three times by centrifugation 
  and redispersed in MQ. 
  
To form a protrusion,
  the CPSAA spheres were swollen with St, heated and polymerized,
  in line with 
  Sheu \textit{et al.}, Mock \textit{et al.} and Kraft \textit{et al.}
  \cite{sheu1990nonsphericalLatex,mock2006synthesisAnisotropicNP,kraft2009colloidalMoleculesBondAngles}.
In a typical experiment,
  the solid mass fraction\textemdash as determined by drying\textemdash
  was brought to 3\textendash6\% with MQ. 
About 5 ml dispersion
  was magnetically stirred
  with a PTFE coated stir bar
  in a glass tube. 
Styrene 
  was added 
  with a swelling ratio $S = m_\textup{St}/m_\textup{s}=3$--7
  with $m_\textup{St}$ and $m_\textup{s}$ 
  being the mass of added St, and the solid mass in the dispersion.
After one to two days stirring,
  the tube was immersed in an 80 \textdegree C oil bath for two hours
  under continuous stirring
  to form a styrene protrusion.
Next, 
  500 \textmu l of an aqueous hydroquinone solution (45 mg/50 ml) 
  and 5 mg AIBN in 250 \textmu l St 
  were added,
  and the tube was immersed in the 80 \textdegree C oil bath for 24 h
  to polymerize the protrusion.
Finally,
  a milky white dispersion 
  was obtained.
A mm-sized solid white aggregate
  was often found,
  which could easily be removed.

Hydrophobic moieties are 
  covalently linked to carbodiimide activated carboxylic groups on CPSAA particles,
  a method adapted from 
  Marchand-Brynaert \textit{et al.}
  \cite{MarchandBrynaert1995surfacefunctionalization}.
In a typical synthesis,
  a 0.1 M MES buffer (1.95 g in 100 ml MQ), 
  and a 0.2 M phosphate buffer (6.72 g Na$_2$HPO$_4$ and 0.41 g NaH$_2$PO$_4\cdot2$H$_2$O  in 250 ml MQ) 
  were prepared.
2.5 ml dispersion 
  was centrifuged,
  and 
  the supernatant 
  was replaced 
  by an EDC/MES solution (45 mg EDC quantitatively added with 40 ml MES buffer)
  to activate the AA groups.
The dispersion 
  was tumbled at 60 rpm for one hour,
  and washed by centrifugation at $2.1\times10^3\,g$
  with MES buffer and
  twice with MQ.
The dispersion 
  was again centrifuged,
  and after removal of the supernatant,
  0.028 mmol FlA or tBA
  was quantitatively added with 30 ml phosphate buffer
  to covalently bind FlA or tBA to the activated AA groups.
The tube was wrapped in aluminium foil,
  and after tumbling at 60 rpm overnight,
  the dispersion 
  was washed
  three times with phosphate buffer,
  once with MES solution,
  and five times with MQ.
Finally,
  a milky white dispersion was obtained.
Covalent linkage of fluoresceinamine
  was verified 
  by varying the fluoresceinamine coupling method 
  and studying the resulting washed particles
  using fluorescence microscopy (Extended Data Fig. \ref{exfgr:Fluorescence}).
The preparation of non-deformable, fluoresceinamine functionalized snowman-like particles 
  was previously described by
  van Ravensteijn \textit{et al.}\cite{ravensteijn2013chemicallyAnisotropicColloids}.

\paragraph{Transmission Electron Microscopy Analysis}
\label{sup:tem}
Transmission electron microscopy (TEM) images 
  were taken with a Philips TECNAI 10 at 100 kV and 16-bit.
Samples were prepared
  by drying a diluted dispersion droplet
  on a polymer coated copper grid
  under illumination with a heat lamp.
Image levels were linearly rescaled 
  using ImageMagick,
  so that 99.9\% of all values were between the lower and upper level thresholds.

Fig. 1a was obtained at 9.7 nm/px and 1b at 0.95 nm/px.
Fig. 2d--e were obtained at 3.5 nm/px, 2f at 6.8 nm/px, and the insets in 2d--f at 0.95 nm/px.
Extended Data Fig. 2b and 2d were obtained
at 4.9 nm/px, and 2c and 2e at 0.95 nm/px.
Extended Data Fig. 5a--d were obtained at 0.9 nm/px.

Particle sizes 
  were measured using Gaussian ring transformations in Wolfram Mathematica 10.
For spherical particles,
  a gradient transform was 
  computed using discrete derivatives of a Gaussian.
Circles were detected
  by iteratively convolving with Gaussian rings,
  and finding the maxima.

\paragraph{Scanning Electron Microscopy}
\label{sup:sem}
By freeze-drying or sintering,
  self-assembled structures
  could be preserved upon drying
  and studied 
  using scanning electron microscopy.
For freeze-drying,
  1 \textmu l dispersion 
  was brought on a polymer coated copper grid.
The grid 
  was vitrified 
  in liquid nitrogen and mounted on a cryotransfer unit
  which was brought under vacuum.
Temperature 
  was increased to -90 \textdegree C at 5 \textdegree C/min 
  and kept constant for about six hours
  to allow the water to sublimate.
  
For sintering,
  the sample was heated above the glass transition temperature of polystyrene at about 100 \textdegree C.
First, the dispersion was centrifuged 
  and after redispersion in 1:1 glycerol:water,
  immersed in an 105 \textdegree C oil bath for 30 minutes.
The dispersion was washed three times by centrifugation with MQ
  and 1 \textmu l 
  was brought on a polymer coated copper grid.
After drying, 
  the sample was coated with a $\sim6$ nm platinum layer.
  
Both samples were studied 
  with a FEI XL30 FEG operated at 5-10 kV,
  and images were obtained at 8-bit.
Image levels were linearly rescaled using ImageMagick,
  from the value of the pixel with the lowest intensity to the brightest pixel.
Fig. 2a was obtained at 6.5 nm/px.
Extended Data Fig. 3c was obtained at 3.5 nm/px, d at
1.9 nm/px, e at 3.5 nm/px, h at 11 nm/px, i at 6.9 nm/px and j at 11 nm/px.

\paragraph{Optical Microscopy}
\label{sup:om}
Bright field, fluorescence and reflected light microscopy
  images were captured with a Nikon Ti-E Inverted Microscope unless stated otherwise.
The Nikon Ti-E was 
  operated 
  with a Nikon TIRF $100\times/1.49$ objective,
  intermediate magnification of $1.5\times$,
  and a Hamamatsu ORCA Flash camera at 16-bit.
For reflected light microscopy,
  a Nikon Intensilight C-HGFl light source was used
  with a Nikon D-FLE filter block. 
For fluorescence microscopy,
  the same light source was used
  with a Semrock FITC-3540C filter block.
The bright field microscopy images
  in Fig. \ref{fgr:Cavities}k--n
  were obtained with a 
  Nikon Ti-U inverted Microscope 
  with a Nikon Plan Apo VC $100\times/1.40$ objective,
  intermediate magnification of $1.5\times$,
  and a Lumenera Infinity X camera at 8-bit.
The bright field microscopy images in
  Fig. \ref{fgr:Cavities}r--s,u--v, Extended Data Fig. \ref{exfgr:Capillary}i--v and Extended Data Fig.  \ref{exfgr:Sediment}
  were obtained 
  with a Nikon Eclipse LV100POL microscope
  with its focal plane parallel to the gravitational field,
  a Nikon Plan Fluor ELWD $40\times/0.6$ objective,
  and a QImaging MicroPublisher 5.0 camera at 8-bit. 
Finally,
  confocal microscopy images
  were captured with a Nikon TE2000-U,
  with a Nikon Plan Apo $100\times/1.40$ objective and
  a 488 nm laser and
  a 590 nm detector at 12-bit.
For images obtained with the TiE and the TE2000-U,
  bitmaps 
  were extracted from the microscopy files
  using bfconvert 5.1.7 (Open Microscopy Environment).
Furthermore, 
  image levels 
  were linearly rescaled using ImageMagick.

Fig. 1c--d were obtained with the TiE in reflected light mode at 40 nm/px and image levels were linearly rescaled from the value of the darkest to the brightest pixel.
Fig. 2b was obtained with the TiE in reflected light mode at 43 nm/px, and levels were linearly rescaled from the value of the darkest pixel to the brightest pixel.
Fig. 2c was obtained with the TE2000-U in confocal fluorescence mode at 35.72 nm/px, and levels were linearly rescaled from zero to the value of the brightest pixel. Fluorescein sodium salt was added to the water phase, and the image was false coloured in green.
Fig. 2g--j were obtained with the TiE in bright field mode at 43 nm/px, and levels were linearly rescaled from zero to the value of the brightest pixel.
Fig. 3b--e were obtained with the TiE in bright field mode at 40 nm/px, and levels were linearly rescaled from zero to the value of the brightest pixel.
Fig. 3g--j were obtained with the TiE in reflected light mode at 43 nm/px, and levels were linearly rescaled from the value of the darkest pixel to the brightest pixel.
Fig. 3k--n were obtained with the TiU in reflect light mode at 29 nm/px, and levels were linearly rescaled from the value of the darkest pixel to the brightest pixel.
Fig. 3r--s,u--v were obtained with the LV100POL at 86 nm/px. For s and v, levels were linearly rescaled from the value of the darkest pixel to the brightest pixel, and these thresholds were also used for r and u.
Extended Data Fig. 2f--g were obtained with the TiE in fluorescence mode at 43 nm/px, and levels were linearly rescaled from zero to the value of the brightest pixel.
Extended Data Fig. 4a was obtained with the TiE in bright field mode at 43 nm/px, and levels were linearly rescaled from zero to the value of the brightest pixel.
Extended Data Fig. 4e--h were obtained with the TiE in bright field mode at 40 nm/px, and levels were linearly rescaled from zero to the value of the brightest pixel.
Extended Data Fig. 4i--v were obtained with the LV100POL at 86 nm/px. For m--o and t--v, levels were linearly rescaled from the value of the darkest pixel to the brightest pixel, and these thresholds were also used for i--k and q--s. For l and p, levels were linearly rescaled from zero to the value of the brightest pixel.
Extended Data Fig. 5e--h were obtained with the TiE in bright field mode at 43 nm/px, and levels were linearly rescaled from zero to the value of the brightest pixel.
The images in Extended Data Fig. 6 were obtained with the TiE in bright field mode at 40 nm/px, and levels were linearly rescaled from zero to the value of the brightest pixel.
The images in Extended Data Fig. 7 were obtained with the LV100POL at 86 nm/px. Image levels were linearly rescaled using the thresholds from Fig. 3s and Extended Data Fig. 4m--o.
Extended Data Fig. 10a--d were obtained with the TiE in fluorescence mode at 43 nm/px, levels were linearly rescaled from zero to the value of the brightest pixel, and the images were false coloured in green.

Typically, a sample was prepared
  by bringing 0.5--2 \textmu l dispersion 
  between a microscope slide (Menzel-Gl\"aser),
  and a $\#1.5$ cover slip (Menzel-Gl\"aser)
  with two $\#0$ cover glasses (VWR) as spacers.
Before use, the slides were cleaned with MQ, ethanol and Kimtech precision wipes,
  and cells were sealed with glue (Norland NOA81, after UV curing)
  or scotch tape.
Evaporating droplets between two glass slides  (Fig. \ref{fgr:Cavities}f--j)
  were studied in cells as described above, 
  but without sealing the sides of the cell.
Evaporating droplets on a glass slide  (Fig. \ref{fgr:Cavities}a--e, Extended Data Fig. \ref{exfgr:Capillary}e--h and Extended Data Fig. \ref{exfgr:ContactLine}),
  on the other hand,
  were studied by bringing 0.5 \textmu l dispersion  
  on a cleaned $\#1.5$ cover slip (Menzel-Gl\"aser).
Droplets evaporated spontaneously
  and for each dispersion, 
  four times (twice for non-deformable particles) a time series,
  was obtained of 89 \textmu m of the contact line (Extended Data Fig. \ref{exfgr:ContactLine}). 
High concentration samples (Fig. \ref{fgr:Cavities}k--n, Supplementary Video 5)
  were studied 
  after sedimentation 
  in the gravitational field
  in a similar cell as described above.
Thin cells as in Fig. \ref{fgr:Cavities}q--v and Extended Data Fig. \ref{exfgr:Capillary}i--v and Extended Data Fig. \ref{exfgr:Sediment}
  were prepared in a capillary and sealed with glue,
  while preventing contact between uncured glue and the dispersion.
A $0.020\times0.200\times50$ mm$^3$ capillary (VitroCom 5002-050, Kimtech cleaned) 
  was filled half with dispersion. 
Next,
  the capillary was pressed on a microscopy slide (Menzel-Gl\"aser) with tweezers and a foam cushion,
  and nitrogen gas was blown from the filling side to push the dispersion to the middle of the tube.
While blowing, 
  the other side was sealed with a glue droplet (Norrland NOA81) to prevent the dispersion from flowing back. 
Finally, 
  the other side was sealed with glue,
  the glue was cured with UV light,
  and the cells were centrifuged in centrifuge tubes (VWR SuperClear).

Interparticle distance distributions 
  were obtained 
  by analysing reflected light microscopy time series
  with Mathematica. 
For each frame,
  the gradient transform was computed using discrete derivatives of a Gaussian, 
  and the gradients were circle transformed by convolving with a circle. 
The original image was multiplied pixel per pixel with the circle transform,
  and the local maximums
  were identified as particles.
The histogram for particles in the cavity phase  (Fig. \ref{fgr:Cavities}o, blue)
  was normalized with a fitted function through the distribution of $10^7$ distances between random points on a plane with the same size as the microscopy images,
  and the resulting histogram 
  was scaled to 1 at large inter-particle distances. 
The effective pair potential  (Fig. \ref{fgr:Cavities}p)
  was calculated using $w(r_{ij})/k_BT=-\ln g(r_{ij})$,
  where $g(r_{ij})$ is the measured interparticle distance distribution.
The histogram for particles in a microcapsule  (Fig. \ref{fgr:Cavities}p, red)
  was scaled so the maximum has the same height as the maximum of the cavity phase histogram.

\label{sup:particlepermicrocapsule}
The number of particles per microcapsule 
  was estimated from the first peak in the interparticle distance distribution of a microcapsule, $r^*_{ij}=0.62$ \textmu m,
  by assuming a hexagonal orientation,
  and a spherical microcapsule surface,
\begin{equation}
  N_\textup{part}
  = \frac{A_\textup{tot}}{A_\textup{part}}
  = \frac{4\pi R_\textup{m}^2}
    {\frac{\sqrt{3}}{2}{r^*_{ij}}^2} 
\end{equation}
 with $A_\textup{tot}$ the surface area of a microcapsule, $A_\textup{part}$ the surface area per particle and $R_\textup{m}$ the radius of the microcapsules.
For particles with a large lobe with diameter 0.540 \textmu m,
  the radii of 30 microcapsules, $R_\textup{m}= 1.9\pm0.4$ \textmu m, was determined similarly as for transmission electron microscopy.
Finally, the estimated number of particles per microcapsule was 
  $N_\textup{part}=10^2$.

\paragraph{Dynamic Light Scattering} 
Cross-linked poly(styrene-\textit{co}-acrylic acid) spheres 
  were added to 1 mM KOH, 
  and using a Malvern Zetasizer Nano ZS, equipped with an MPT-2 autotitrator, 
  pH was stepwise decreased from 9.8 to 3.4 
  by adding 1 mM HCl.
Cumulant analysis 
  gave the apparent size and polydispersity index.
The influence of electrostatic interactions
  was studied by 
  replacing the solvents with
  1 mM KOH / 9 mM KCl 
  and 1 mM HCl / 9 mM KCl.
The influence of the carboxylic acid groups
  was studied by using cross-linked polystyrene spheres instead of CPSAA spheres.

\paragraph{Simulation model}
\label{sup:sim_model}
We model the colloids as hard-spheres of diameter $\sigma$.
The attractive forces between particles, due to e.g. van der Waals interactions and the hydrophobic (polystyrene) component of the brush, is captured using a square well potential:
\begin{equation}
\label{eq:Ucc}
U_{\rm cc}(R) = \left\{ 
  \begin{array}{l l}
    \infty & \quad R \leq \sigma,\\
    -\varepsilon_{\rm c} & \quad \sigma < R \leq \sigma + \Delta,\\
    0 & \quad R >  \sigma + \Delta,
  \end{array} \right.
\end{equation}
where $\Delta$ is the width of the well and $R$ is the centre-centre distance between the colloids.
We then model the grafted acrylic acid polymers as penetrable hard spheres (PHS) or satellite spheres with diameter $q\sigma$ \cite{asakura1954twoBodyInteractionMacromoleculeSolution}, which can move freely across the surface at a fixed distance $(\sigma + q\sigma)/2$, thereby allowing the brush to adapt its configuration to the environment.
The PHS particles are free to overlap each other, i.e. $U_{\rm PHS-PHS} = 0$, but have a hard-sphere interaction with the colloids:
\begin{equation}
U_{\rm PHS-c}(r) = \left\{
  \begin{array}{l l}
    \infty & \quad  r \leq (\sigma + q\sigma)/2,\\
    0 &\quad r > (\sigma + q\sigma)/2,
  \end{array}\right.
\end{equation}
where $r$ is the centre-centre distance between the colloid and polymer.

To obtain snowman-like particles, we attach a protrusion to each colloid, modelled here as a hard-sphere with radius $r_{\rm p}$, with its centre located at the colloid surface.
The protrusion hydrophobicity is captured by a weak square-well attraction between protrusions with strength $ U_{\rm pp} = -\varepsilon_{\rm p}$ over the range $2r_{\rm p} < R \leq 2r_{\rm p} + \Delta$.
The colloids and protrusions interact through a square-well potential with depth $U_{\rm cp} =- \sqrt{\varepsilon_{\rm c} \varepsilon_{\rm p}}$, i.e. the geometric mean of the interaction strengths, over the range $\sigma/2 + r_{\rm p} < R \leq \sigma/2 + r_{\rm p} + \Delta$.
Lastly, the protrusions have a hard-sphere interaction with the polymers:
\begin{equation}
U_{\rm PHS-p}(r) = \left\{
  \begin{array}{l l}
    \infty & \quad  r \leq r_{\rm p} + q\sigma/2,\\
    0 &\quad r > r_{\rm p} + q\sigma/2.
  \end{array}\right.
\end{equation}

Given a system containing $N$ particles at positions $\textbf{R}^N$, each grafted with $f$ polymers at positions $\textbf{r}^f$, the total energy for the system is then given by
\begin{equation}
\begin{aligned} 
\label{eq:b2sum}
&U(\textbf{R}^N, \textbf{r}^{fN})  = \sum_{i \neq j}^N \Bigl [ U_{\rm cc}(\vec{R}_{ij}) +U_{\rm cp}([\vec{R}_{i} + \vec{u}_i] - \vec{R}_j) \\
&+ U_{\rm cp}(\vec{R}_{i} - [\vec{R}_j+ \vec{u}_j] ) + U_{\rm pp} ( [\vec{R}_{i} + \vec{u}_i]- [\vec{R}_j+ \vec{u}_j] ) \Bigr ]\\
&+ \sum_{i,j}^N \sum_k^f \Bigl [ U_{\rm PHS-c} (\vec{r}_{i,k} - \vec{R}_j) + U_{\rm PHS-p} (\vec{r}_{i,k} - (\vec{R}_j + \vec{u}_j))\Bigr ],
\end{aligned}
\end{equation}
where $\vec{r}_{i,k}$ denotes the position of the $k$-th polymer of the $i$-th particle, and $\vec{u}_{i} = M(\vec{\Omega}_i) \Delta  \vec{u}$ is the vector pointing from the centre of the colloid of the $i$-th particle to the protrusion, with $M(\vec{\Omega}_i)$ the rotation matrix for the orientation $\vec{\Omega}_i$ of the $i$-th nanoparticle, and $\Delta \vec{u}$ the vector in the reference frame from the centre of the nanoparticle to the protrusion \cite{luiken2013anisotropicPolymerNP}.

\paragraph{Simulation details}
\label{sup:sim_details}
We employ Monte Carlo (MC) simulations in the canonical NVT ensemble to study the aggregation behaviour of mutually attractive, anisotropic, deformable  colloids.
We perform two types of simulations: initiating from i) the soluble phase and increasing the interaction strength from $\varepsilon_{\rm c} = 3$ to 9 $k_BT$ in steps of 1 $k_BT$, ii) from a square planar monolayer in a hexagonal packing arrangement with fixed interaction strength $\varepsilon_{\rm c} = 9$ $k_BT$.
We set $N=98$ 
and the box length $L$ is set such that the colloid packing fraction equals $\phi_{\rm c} = 0.001$; periodic boundary conditions apply in all directions.
The protrusions have a radius $r_{\rm p} = 0.35\sigma$, they are weakly attractive relative to the colloids: $\varepsilon_{\rm p}  = \varepsilon_{\rm c}/5$, and are randomly oriented below and above the monolayer in system ii. 
Lastly, the square well width is set to $\Delta= 0.1\sigma$.

We perform $4 \times 10^4$ equilibration MC cycles and another $150 \times 10^4$ production cycles for each step in system i.
Per MC cycle we attempt 50 colloid displacement moves over a fixed maximum distance ($0.25L$), 50 colloid moves with variable maximum displacement such that $25\% < P_{\rm acc} < 40\%$, 50 quaternion rotations of the protrusion and another 50 of the entire nanoparticle, 50 cluster moves \cite{bhattacharyay2008clusterAlgorithm} over a fixed distance ($0.15L$) 
and $50f/2$ quaternion rotations of polymers.
We disable the expensive cluster moves for system ii, allowing us to greatly increase the number of equilibration cycles to $2 \times 10^6$ and the number of production cycles to $10 \times 10^6$.

For all three systems we perform simulations for every combination of functionality $f \in \{ 2,4,6,8,10,12 \}$ and polymer size $q \in \{ 0.20, 0.25, ..., 0.70 \}$, and we repeat all simulations in the absence of protrusions.
We repeat the simulations initiated from the hexagonal monolayer an additional four times, and all resulting data plots and morphologies are averages over these runs.

\paragraph{Simulation data analysis}
\label{sup:sim_analysis}
We evaluate the average number of bonds per particle using the expression
\begin{equation}
\label{eq:Nc}
\langle{N}_{\rm b}\rangle = \frac{1}{N} \sum_{i \neq j}^N \theta(\sigma+\Delta-r_{ij}),
\end{equation}
where $\theta$ is the Heaviside step function.
Here, particles are in contact when the centre-centre distance $r_{ij} =  |\vec{r}_i - \vec{r}_j |$ between their bodies is less than $\sigma + \Delta$.

We define the covered surface fraction
 as the surface which is covered by surface groups divided by the total available surface,
\begin{equation}
\label{eq:Q}
Q \equiv \frac{f A_q}{A_{\rm tot} - A_{\rm ex}},
\end{equation}
where $A_q$ is the surface area covered by a satellite sphere, $A_{\rm tot} = 4\pi ((\sigma+\sigma q)/2)^2$ is the total area of the sphere over which the satellite spheres move, and $A_{\rm ex}$ is the surface area excluded by the presence of the protrusion.

$A_q$ and $A_{\rm ex}$ are the curved surface areas of spherical caps. These areas are given by $A_{\rm cap} = 2\pi R_{\rm cap}h_{\rm cap}$, with $R_{\rm cap}$ the radius of the sphere, and $h_{\rm cap}=R_{\rm cap}-R_{\rm cap}\cos\theta_{\rm cap}$ the height of the cap. In the last equation, $\theta_{\rm cap}$ is the angle between the edge of the cap, the centre of the sphere and the centre of the cap. Inserting this in the equation for $A_{\rm cap}$ gives,
\begin{equation}
  \label{eq:Scap}
  A_{cap} 
  = 2\pi R_{\rm cap}^2 (1-\cos\theta_{\rm cap}).
\end{equation}

For the satellite spheres, 
  $R_{\rm cap}=(\sigma+\sigma q)/2$
  and the law of cosines gives,
  \begin{equation}
    \cos\theta_{\rm cap} =  
    \frac{
      (\frac{\sigma+q\sigma}{2})^2 
      + (\frac{\sigma+q\sigma}{2})^2 
      - (\frac{q\sigma}{2})^2
    }{
      2
      (\frac{\sigma+q\sigma}{2})
      (\frac{\sigma+q\sigma}{2})
    }
    = 1-\frac{q^2}{2(1+q)^2},
  \end{equation}
and for the excluded surface area,
  $R_{\rm cap}=(\sigma+\sigma q)/2$
  and,
  \begin{equation}
  \label{eq:cost}
    \cos\theta_{\rm cap} = 
    \frac{
      (\frac{\sigma}{2})^2 
      + (\frac{\sigma+q\sigma}{2})^2 
      - (\frac{p\sigma+q\sigma}{2})^2
    }{
      2
      (\frac{\sigma}{2})
      (\frac{\sigma+q\sigma}{2}
      )
    }
    = 1-\frac{2pq+p^2}{2+2q},
  \end{equation}
  with $p=2r_p/\sigma$, the dimensionless protrusion diameter.
  
Inserting Eq. \ref{eq:Scap}--\ref{eq:cost} in Eq. \ref{eq:Q} gives,
\begin{equation}
\label{eq:csfSnowmen}
Q=
  \frac{
      f q^2
    }{
      (1+q)(4+4q-2pq-p^2)
    }.
\end{equation}
For particles without protrusions, $p = 0$, this reduces to,
\begin{equation}
Q = \frac{f q^2}{4(1+q)^2}.
\end{equation}

\clearpage
\renewcommand{\figurename}{Extended Data Figure}
\setcounter{figure}{0}

\begin{figure}[!ht]
	\centering
	\includegraphics{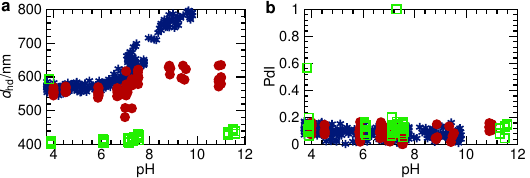}
  \caption{
  \textit{pH-induced structural rearrangements}. 
For poly(styrene-\textit{co}-acrylic acid)  spheres with a transmission electron microscopy diameter, $d=0.530\pm$0.014 \textmu m,
  the apparent hydrodynamic diameter, $d_\textup{hd}$, is measured using dynamic light scattering (a).
At ionic strength $I\approx1$ mM,
  $d_\textup{hd}$ equals $0.79$ \textmu m at pH 10,
  but decreases to 0.57 \textmu m at pH 3 (blue).
Upon screening electrostatic interactions at $I\approx10$ mM (red), 
  or for polystyrene spheres without acrylic acid (green),
  however,
  the measured diameter remains almost constant with pH. 
We conclude that
  at high pH,
  electrostatic repulsion between acrylic acid groups
  triggers the poly(acrylic acid)-rich brush to expand about 0.1 \textmu m
  into the solution.
The measured polydispersity index, PdI,
  stays constants (b), indicating that changing the pH does not induce aggregation.
}
\label{exfgr:DLS}
\end{figure}

\begin{figure}[!ht]
	\centering
	\includegraphics{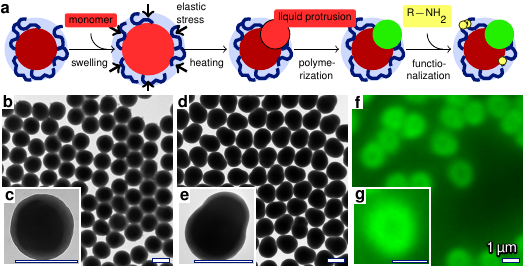}
  \caption{
  \textit{Synthesis.} Schematic outline (a) and microscopy images (b--g) of the synthesis of mutually attractive, anisotropic, deformable particles. Poly(styrene-\textit{co}-acrylic acid) spheres (b--c, transmission electron microscopy) with a hydrophobic core (a, red) and a deformable brush (a, blue) are swollen with monomer, heated, and polymerized, resulting in snowman-like particles (d--e, transmission electron microscopy) with a deformable lobe and a rigid protrusion (a, green). Hydrophobic molecules (a, yellow) are covalently linked to the acrylic acid groups, resulting in  fluorescent particles when fluoresceinamine is used (f--g, fluorescence microscopy).
}
\label{exfgr:Synthesis}
\end{figure}

\begin{figure}[!ht]
	\centering
	\includegraphics{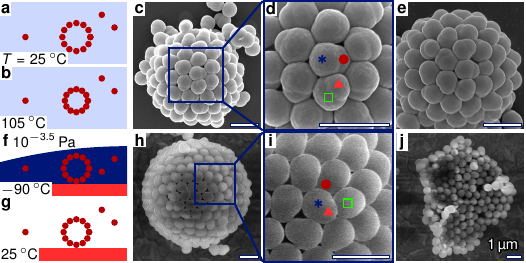}
  \caption{
\textit{Scanning electron microscopy images of self-assembled microcapsules.} 
To prevent disintegration upon drying, microcapsules are studied after sintering (a--e) or freeze-drying (f--j). During sintering, the solvent (light blue) is heated in order to partly merge the particles (red) (a--b). 
During freeze-drying, vitrified water (dark blue) is sublimated under vacuum (f--g). Particles in the microcapsules have six (d,i blue) or five (green) neighbours, and both protrusions that point slightly inwards (red) and outwards (light red) are found. Besides microcapsules, also planar monolayers (j) are observed. 
}
\label{exfgr:SEM}
\end{figure}

\begin{figure}[!ht]
	\centering
	\includegraphics{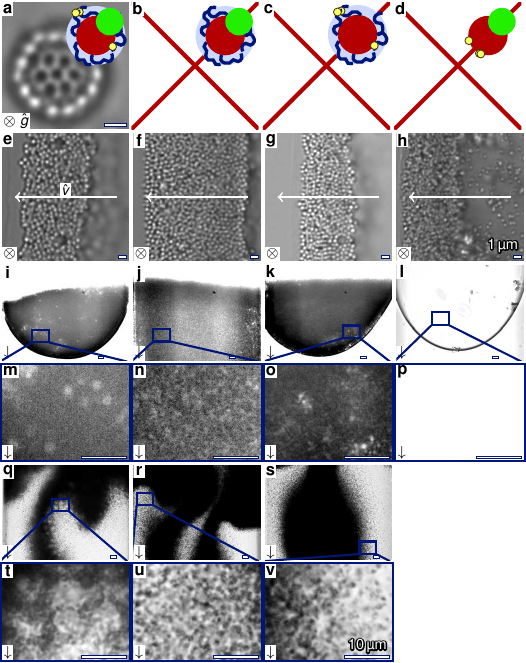}
  \caption{
\textit{Formation of microcapsules and cavities upon varying the complexity of the particles.} 
The complexity of mutually attractive, anisotropic, deformable particles (a) that are deformable (blue), anisotropic (green) and functionalized with mutually attractive groups (yellow) is varied resulting in non-functionalized particles (b), isotropic particles (c) and non-deformable particles (d). Microcapsules are only found in the first case (a). Furthermore, all particles are studied using bright field microscopy at the edge of an evaporating droplet (e--h), in a sediment after centrifugation (i--p), and upon diluting the sediment (q--v). The entire images of e--h can be found in Extended Data Fig. \ref{exfgr:ContactLine} and magnifications of i--k can be found in Extended Data Fig. \ref{exfgr:Sediment}b--d. The arrows indicate the directions of the particle flow, $\hat{v}$, and the gravitational field $\hat{g}$.
}
\label{exfgr:Capillary}
\end{figure}

\begin{figure}[!ht]
	\centering
	\includegraphics{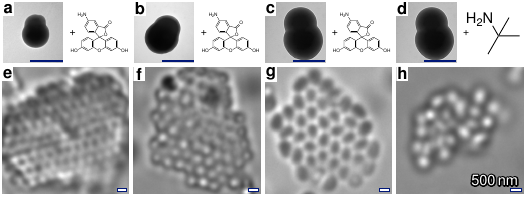}
  \caption{
  \textit{Monolayer sheets.} a) For mutually attractive, anisotropic, deformable particles with varying sizes (a-c), not only hollow microcapsules (Fig. \ref{fgr:Microcapsules}), but also two-dimensional hexagonal planar monolayers (e-g) are observed using bright field microscopy. Both  fluoresceinamine (a--c,e--g) and \textit{tert}-butylamine (d,h) are used as hydrophobic moieties and for both moieties monolayers are observed.
}
\label{exfgr:Sheets}
\end{figure}

\begin{figure}[!ht]
	\hspace{-2cm}%
	\includegraphics{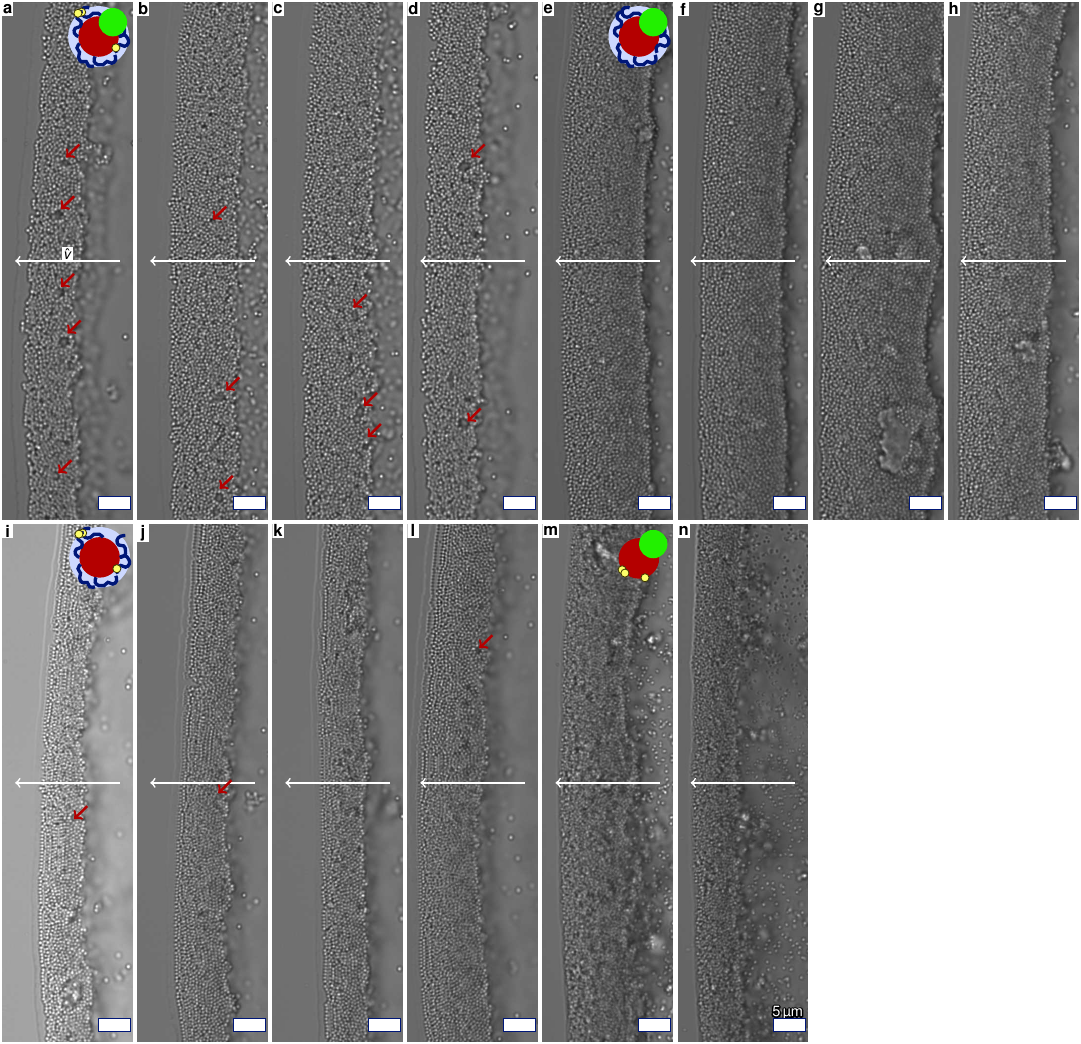}
  \caption{
  \textit{Formation of cavities at the contact line.} The complexity of mutually attractive, anisotropic, deformable particles (a--d) is varied resulting in non-functionalized particles (e--h), isotropic particles (i--l) and non-deformable particles (m--n). For each type, four times (twice for non-deformable particles) a part of the contact line of an evaporating droplet is studied. Many cavities are found for mutually attractive, anisotropic, deformable particles (red arrows), while for isotropic particles much less cavities are found, and the other particles did not show any cavities. Crops of these images can be found in Extended Data Fig. \ref{exfgr:Capillary}. The white arrows indicate the direction of the particle flow, $\hat{v}$.
}
\label{exfgr:ContactLine}
\end{figure}

\begin{figure}[!ht]
	\hspace{-2cm}%
	\includegraphics{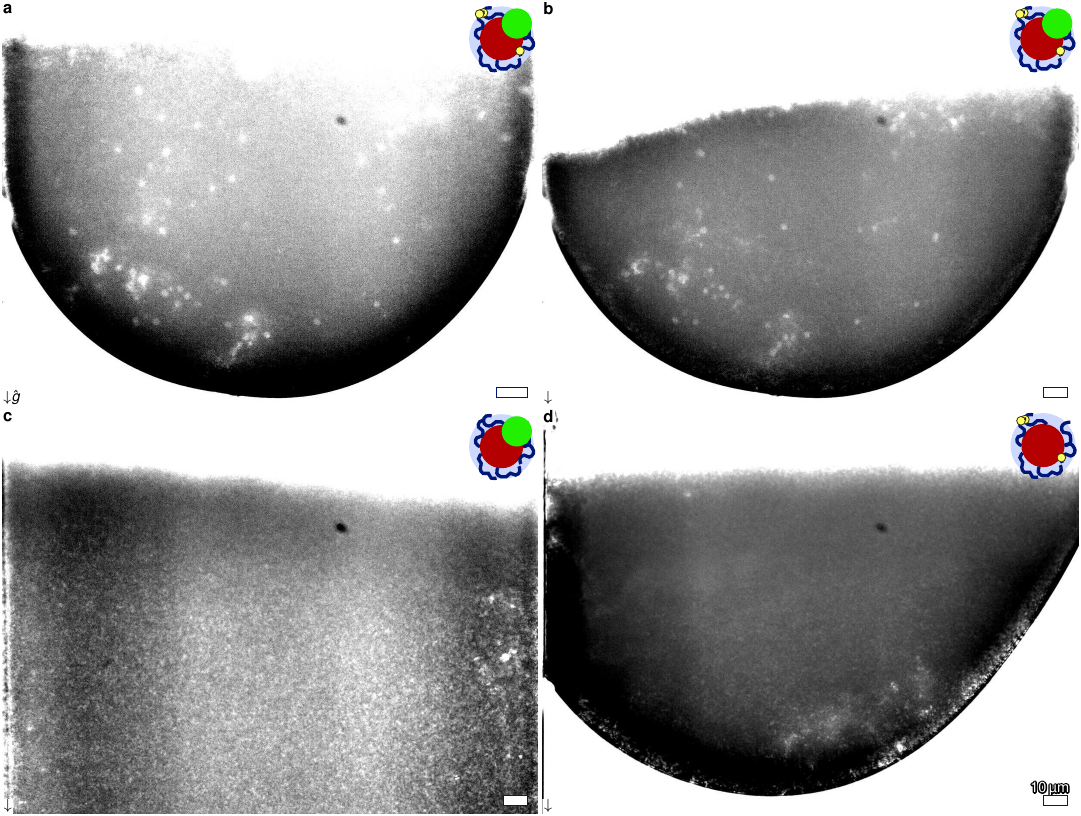}
  \caption{
  \textit{Centrifuged sediments.} Magnifications of the bright field microscopy images Fig. \ref{fgr:Cavities}r and Extended Data Fig. \ref{exfgr:Capillary}i--k. For mutually attractive, anisotropic, deformable particles, spherical cavities are observed in the sediment (a--b), while the sediments of similar non-functionalized and isotropic particles show no (c) and less (d) cavities.
}
\label{exfgr:Sediment}
\end{figure}

\begin{figure}[!ht]
	\centering
	\includegraphics{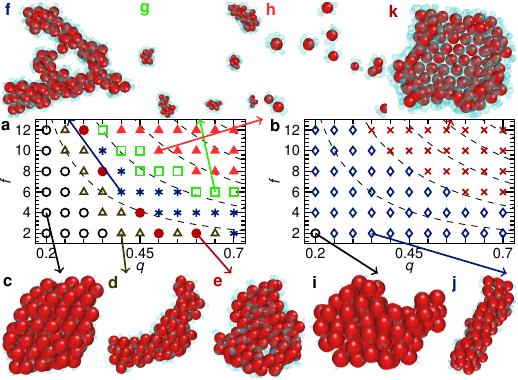}
  \caption{
  \textit{Clusters of isotropic particles.} a-b) Morphology diagrams of mutually attractive, isotropic, deformable particles as a function of the dimensionless diameter of the satellite spheres, $q$, and the number of satellite spheres, $f$, and c--k) representative snapshots with cores (red) and satellite spheres (blue). a) When unbound particles are used as the initial configuration and $q$ and $f$ are increased, 
compact (c,\SmallCircle), 
cylindrical (\textcolor{xBrown}{d,\SmallTriangleUp}),
flattened (\textcolor{xRed}{e,\FilledSmallCircle}), 
rod-like (\textcolor{xBlue}{f,$\ast$})
and finite-size (\textcolor{xGreen}{g,\SmallSquare}) clusters as well as 
unbound particles (\textcolor{xLightred}{h,\FilledSmallTriangleUp}) are found. 
b) When the initial configuration is a hexagonal monolayer,
compact clusters (i, \SmallCircle), 
bilayers (\textcolor{xBlue}{j,\SmallDiamondshape}) 
and monolayers (\textcolor{xRed}{k,\SmallCross}) are observed.
The transitions between different morphologies are parallel to isolines for the covered surface fraction, $Q=0.1$ to 0.5 (dashed).
}
\label{exfgr:Morphology}
\end{figure}

\begin{figure}[!ht]
	\centering
	\includegraphics{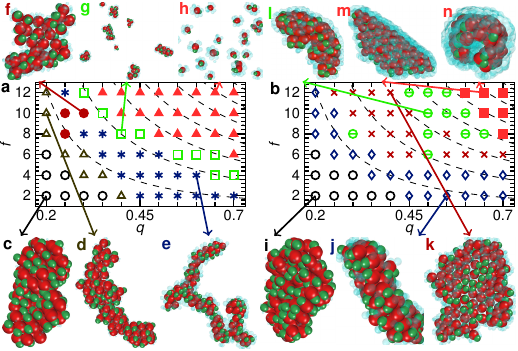}
  \caption{
\textit{Clusters of anisotropic particles.} a-b) Morphology diagrams of clusters of mutually attractive, anisotropic, deformable particles as a function of the dimensionless diameter of the satellite spheres, $q$, and the number of satellite spheres, $f$, and c--n) representative snapshots with cores (red), protrusions (green) and satellite spheres (blue). 
When unbound particles are used as the initial configuration (a), increasing $f$ and $q$ results in 
compact (c,\SmallCircle), 
cylindrical (\textcolor{xBrown}{d,\SmallTriangleUp}), 
  rod-like (\textcolor{xBlue}{e,$\ast$}), 
  flattened (\textcolor{xRed}{f,\FilledSmallCircle}) and 
  finite-size (\textcolor{xGreen}{g,\SmallSquare}) clusters as well as 
  unbound particles (\textcolor{xLightred}{h,\FilledSmallTriangleUp}). 
  When the initial configuration is a hexagonal monolayer (b), 
  compact clusters (i,\SmallCircle), 
  bilayers (\textcolor{xBlue}{j,\SmallDiamondshape}), 
  planar monolayers (\textcolor{xRed}{k,\SmallCross}), 
  curved monolayers with in-plane protrusions (\textcolor{xGreen}{l,\rlap{\HBar}\SmallCircle}) and 
  curved monolayers with out-of-plane protrusions (\textcolor{xLightred}{m--n,\FilledSmallSquare}) are found. 
  The transitions between different morphologies are parallel to isolines for the covered surface fraction, $Q=0.1$ to 0.7.
}
\label{exfgr:MorphologySnowmen}
\end{figure}

\begin{figure}[!ht]
	\centering
	\includegraphics{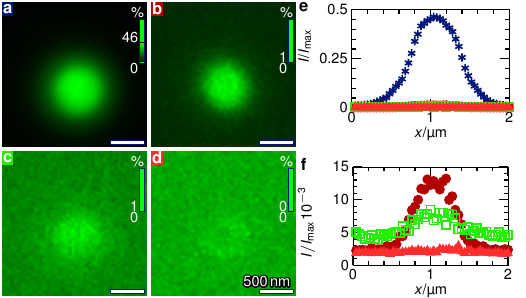}
  \caption{
  \textit{Functionalized CPSAA spheres.} (a-d) Fluorescence microscopy images for variations on the fluoresceinamine coupling method, and (e--f) normalized fluorescence intensity, $I/I_\textup{max}$, as a function of the distance, $x$, on the horizontal line through the fluorescence maximum.  Poly(styrene-\textit{co}-acrylic acid) spheres activated and functionalized as described in Methods (a,e--f blue). For similarly activated and functionalized polystyrene spheres (b,e--f red). For CPSAA without \textit{N}-(3-Dimethylaminopropyl)-\textit{N}'-ethylcarbodiimide hydrochloride addition (c,e--f green). For CPSAA without fluoresceinamine addition (d,e--f light red). The vertical bars indicate the image level thresholds.
}
\label{exfgr:Fluorescence}
\end{figure}

\end{document}